\documentstyle[psbox,graphicx,rotating]{WORKSHOP}

\def\ros{{\em ROSAT }}
\def\cha{{\em CHANDRA }}
\def\xmm{{\em XMM-Newton }}
\def\iras{{IRAS~13224--3809 }}

\begin{document}

% TITLE OF THE PAPER
%  If the title is too long for a single line, you can split it 
%  by putting two backslashes. 
%  You might want to put the subtitle. Then it should be inserted 
%  within {\large\sf  }.
%  e.g.:  
%     \title{ Too Long Title \\ for one line \\
%     {\large\sf Subtitle} }

\title{\cha observation of the narrow-line Seyfert 1 galaxy IRAS~13224-3809 \\ 
%{\large\sf  -- Brief Instructions for Users of the `WORKSHOP' Style File --}
}

\author{
F. Pfefferkorn, Th. Boller, V. Burwitz and P. Predehl
\\[12pt]  % TO BE SPACED WITH ONE LINE
%
% INSTITUTES OF AUTHORS
Max-Planck-Institute for Extraterrestrical Physics, Giessenbachstr.,
85748 Garching, Germany\\
%$^2$  Institute and its address of author3 \\
%
% please put the first author's initial and e-mail address below
{\it E-mail(FP): pfefferk@mpe.mpg.de} 
%            \_ Initial \
%                                      \_ E-mail address
}

\abst{The narrow-line Seyfert 1 galaxy (NLS1) \iras has been observed with the \cha
High Resolution Camera (HRC-I) for 12 ksec on February, 2, 2000. The source was
proposed for \cha observations to precisely determine the X-ray centroid
position and to investigate the timing properties of the most X-ray variable
Seyfert galaxy. The position derived from \cha confirms that the X-ray
emission is associated with IRAS~13224--3809.
The \cha HRC-I light curve shows indications for a possible presence of a
quasi-periodic oscillation. The strongest signal is found at 2500 sec.
Accretion disk instabilities may provide a plausible explanation for the
quasi-periodic oscillations.}

\kword{galaxies: active --- galaxies: individual: IRAS 13224--3809 --- galaxies: Seyfert --- X-rays: galaxies}
 
\maketitle
\thispagestyle{empty}

\section{Introduction}
The narrow-line Seyfert 1 galaxy (NLS1) \iras was the first galaxy
proposed for monitoring observations with the \ros High Resolution Imager
(HRI) (Boller et al., 1997). The most important observational fact is the
detection of multiple strong flaring events during the 30 day observations. The 
authors found five giant-amplitude variations with the most extreme variability
of a factor of 57 in 2 days. In addition, other \ros observations were
performed, mostly with the HRI detector during 1992 and 1998.
To measure the precise X-ray position, \iras has been proposed for
an observation with the \cha satellite.\\

\section{Data reduction}
\label{observation}

\iras has been observed with the Chandra High Resolution Camera (HRC-I) on February,
2, 2000 between 15:16:28.59 UT and 18:41:04.25 UT with an effective exposure time
of 12271.6 sec.

Inspection of  the image with the source constructed from the reprocessed level 2 
photon event file (hrcf00328N002\_evt2.fits) indicates that the 
aspect reconstruction was not yet perfect as the  FWHM  of the source
is $\sim$ 1.04 arcsec which is larger than that expected from
the instrument. A detailed analysis revealed that the 
source moved around the mean source position 
by approximately $\pm$ 0.5 arcsec with the dither period.
This analysis was performed by computing the centroid in x, y, t
for 100 consecutive events that lie within a circle of 2 arcsec around 
the mean source position. This 100 event window is shifted always 
by one event giving a smooth curve which shows the displacement of the
centroid versus time giving us a correction vector. 
This vector is then used to correct the position of the individual
photon events. The data reduction, correction, and extraction was performed
using routines written in IDL. Using the correction vector we obtain a more
realistic FWHM of $\sim$ 0.64 arcsec for IRAS~13224--3809.  
This value is still larger than the expected value for 
the instrument indicating that extended X-ray emission is present 
around IRAS~13224--3809. A maximum likelyhood source detection usinig
CELLDETECT (from the CIAO data analysis package) performed on the binned
image also indicates that the source FWHM is greater than that of the
instrumental PSF (PSF-ratio=1.43187).

After checking the radial profile of the source we chose an   
extraction radius of 9 arcsec around the source so that also
the photons from the diffuse emission could be collected.
The background selection was performed by selecting photons in 8 regions, 
each with a radius of 9 arcsec. The regions are located 20 arcsec from the
source.

During the observation the total number of events, especially all
background, rose by a factor $\sim$ 3-4 over a time span of $\sim$ 2000 sec.
Higher than normal solar activity is the most probable cause for this increase,
which leads to a leakage of source counts based on the dead-time of the
instrument. The ratio of detected to lost photons versus time is given in
deadtime correction file. This file was then used to correct lower count rate
of the source during times with high background count rates. Only the photons
lying in the goodtime intervals were used.

\section{The \cha position of IRAS~13224--3809}
\label{position}

\begin{figure}[h!]
\centering
%\psbox[xsize=0.4#1,ysize=0.2#1,rotate=r]
\psbox[xsize=8.0cm]{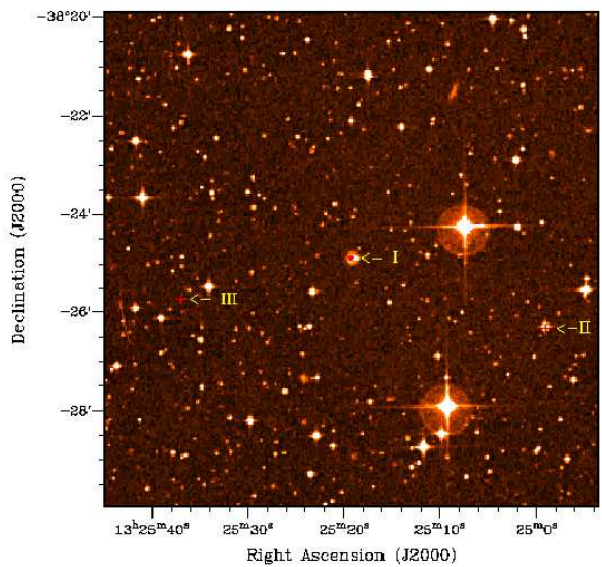}
\caption{Optical image (taken from DSS) overlaid with \cha X-ray contour map of IRAS~13224--3809 and marked
\cha X-ray position of the sources GSC 7787 01660 (II) and RX J132537--3825 (III).}
\label{fig_overlay}
\end{figure}

With the \cha satellite one can determine the position of the X-ray emission
with an unprecedented precision.
In Fig. \ref{fig_overlay}, we show an optical image taken from the Palomar
Digitized Sky Survey (DSS) overlaid with a \cha X-ray contour map 
of IRAS~13224--3809.

\begin{figure}[ht]
\centering
%\psbox[xsize=0.4#1,ysize=0.2#1,rotate=r]
\psbox[xsize=6.0cm]{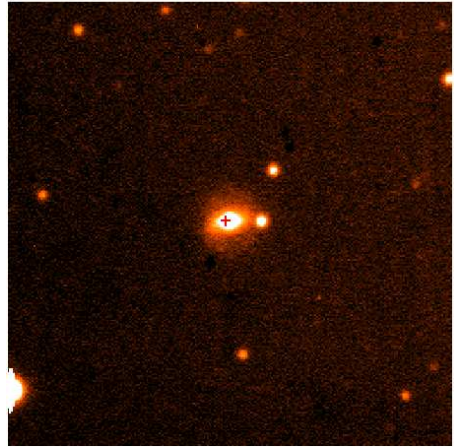}
\caption{\cha X-ray positon marked with red cross in a deeper optical image
(subrange of Fig. \ref{fig_overlay}) with higher resolution observed at the
ESO-Danish 1.54m telescope.}  
\label{fig_optical}
\end{figure}

The \cha X-ray positions of two other sources, GSC 7787 01660
(II) and RX J132537--3825 (III), are marked by a cross. Note that no optical
counterpart is found in optical catalogues for the source RX J132537-3825
(III). For the source detection we have used CELLDETECT from the \cha  CIAO
data analysis package. The centroid position for \iras in the \cha image,
computed with CELLDETECT, is $\alpha_{2000} = 13^h25'19.41''$,
$\delta_{2000} = -38^{\circ}24'52.85''$.
The coordinates of \iras and the two other X-ray sources are listed in table
\ref{coordinates}.
Fig. \ref{fig_optical} shows an higher resolution optical image with the \cha
postion marked by a red cross. The agreement between the X-ray positions and
optical positions obtained from the GSC catalogue is within 2 arcsec. 

\begin{table*}[ht!]
\caption{\bf\underline{Coordinates:} \rm\small{X-ray (using
CELLDETECT) and optical coordinates (taken from {\rm HST} GSC catalogue) of
IRAS~13224--3809, GSC 7787 01660 (RX J132459--3826) and RX J132537--3825.}} 
\label{coordinates}
%\vspace{0.5cm}
%\scriptsize
\begin{center}
%\begin{flushleft}
\begin{tabular}{llccccrr} \hline\hline
name & & \multicolumn{2}{c}{\cha position} &\multicolumn{2}{c}{optical
position} &$\Delta\alpha$ & $\Delta\delta$\\
& & $\alpha_{2000}$ & $\delta_{2000}\quad$ & $\alpha_{2000}$ &
$\delta_{2000}\quad$ & & \\
& & $\rm [h]\,[m]\;[s]\;\;$ & $[^\circ]\;[']\,['']$ &  $\rm
[h]\,[m]\;[s]\;\;$ & $[^\circ]\;[']\,['']$ & $['']$ & $[''] $ \\ \hline
IRAS~13224--3809 & (I) & 13 25 19.41 & -38 24 52.85 & 13 25 19.28 & -38 24 53.48 & 1.95 & -0.63 \\
GSC 7787 01660 & (II) & 13 24 59.20 & -38 26 18.53 & 13 24 59.20 & -38 26 17.18 & 0.0 & 1.35 \\ 
RX J132537--3825 & (III) & 13 25 37.09 & -38 25 42.66 & - & - & - & - \\ \hline\hline  
\end{tabular}
%\end{flushleft}
\end{center}
\end{table*}

\section{Timing properties of IRAS~13224--3809}
\label{timing prop}

In this section we discuss the properties of the light curve of IRAS~13224--3809,
including the indications for a quasi-periodic signal and the tests for periodicity.

\subsection{Lightcurve of IRAS~13224--3809}
\label{light}

\begin{figure}[h!]
\vspace{-10mm}
\centering
%\psbox[xsize=0.4#1,ysize=0.2#1,rotate=r]
\psbox[xsize=10.0cm,rotate=r]{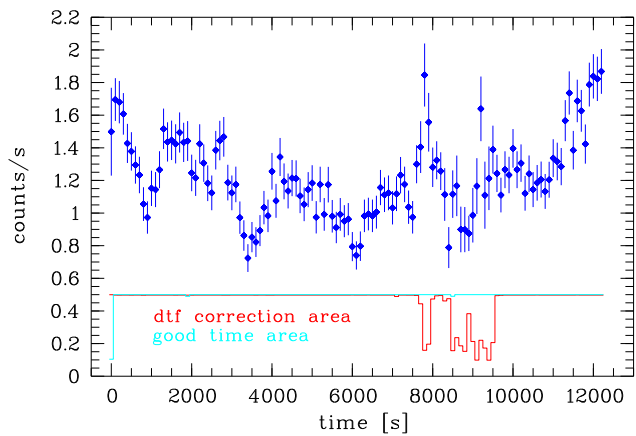}
\caption{Light curve of IRAS~13224--3809 combined with deadtime correction file (dtf) and goodtime
correction file of the second processed data (correction files are shifted down
by factor of $\frac{1}{2}$). Note the larger error in the 2000 sec region of dtf
correction - for description see section \ref{observation}. The light curve
indicate a quasi periodicity with a period of 2500 s.}
\label{fig_light}
\end{figure}

\begin{figure}[h!]
\centering
%\psbox[xsize=0.4#1,ysize=0.2#1,rotate=r]
\psbox[xsize=9.0cm,rotate=r]{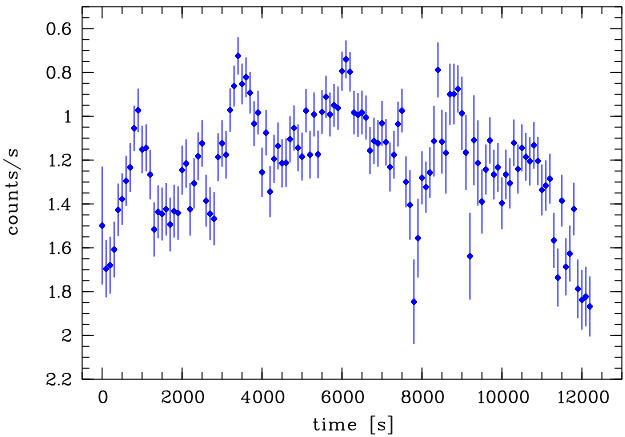}
\caption{Light curve with inverse count rate axis of IRAS~13224--3809 illustrate the sharp
and clearly separated peaks. The light curve indicate a quasi periodicity with
a period of 2500 s.} 
\label{fig_inv_light}
\end{figure}

In Fig. \ref{fig_light} we show the \cha light curve of IRAS~13224--3809.
The reason for the lack of flaring events compared to the \ros observations
in 1996 (Boller et al. 1997) is probably due the short exposure time (12 ksec) of
the observation. However, the light curve exhibits flux variations with the
strongest flux change by a factor 2 within 600 sec.
The efficiency calculation shows that the change of luminosity over time
($\rm \Delta L / \Delta t = 3.2 \cdot 10^{41} erg s^{-2}$) exceeds the maximum
allowed value by accretion onto a Schwarzschild black hole.  This fact points
to the presence of a Kerr black hole in IRAS~13224--3809. The 'peak emission' of
the light curve is not well defined and flattened, however four well-defined
count rate minima are clearly visible and separated by 2500 sec. This fact and
the shape of the count rate variations seems to be an indication for a
quasi-periodic signal in IRAS~13224--3809. In Fig. \ref{fig_inv_light}, the
light curve with an inverted count rate axis illustrates the sharp and clearly
separated peaks at 900, 3400, 6000, 8700 and 10800 seconds. 

The mean count rate of IRAS~13224--3809,
obtained from the light curve, is $\rm cps = 1.222\pm0.010\;
counts\;s^{-1}$. The unabsorbed flux in the 0.1 - 10.0 keV energy range
computed with W3PIMMS\footnote{http://heasarc.gsfc.nasa.gov/Tools/w3pimms.html}
using a power law and black body model with the spectral
parameters; hydrogen column density $\rm N_{H} = (0.87\pm0.05) \cdot
10^{21}cm^{-2}$ (Boller et al., 1996), photon index $\rm \Gamma = 2.47$,
temperature $\rm kT = 0.121\; keV$  (Brandt, private communication) and the
relative normalisation $n = 0.2181$  at 1.0 keV; results in $\rm F_X =
(4.310\pm0.035) \cdot 10^{-11} erg\;cm^{-2}\;s^{-1}$ corresponding to the X-ray
luminosity $\rm L_X = (3.880\pm0.036) \cdot 10^{44} erg\;s^{-1}$ (for $z = 0.0667$,
$\rm q_0 = \frac{1}{2}$ and $\rm H_0 = 75\;km\;s^{-1}Mpc^{-1}$).

\subsection{Tests of periodicity - $\chi^2$, FFT, Lomb - Scargle}

\begin{figure}[h!]
\centering
%\psbox[xsize=0.4#1,ysize=0.2#1,rotate=r]
\psbox[xsize=9.0cm,rotate=r]{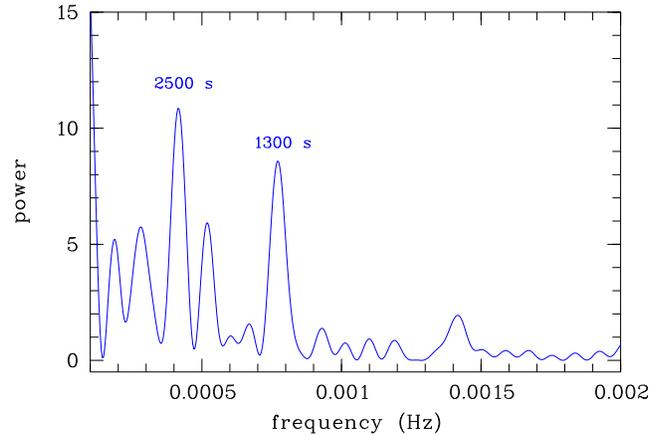}
\caption{Lomb scargle test for periodicity in the 0.2-10 keV energy range.}
\label{fig_lomb}
\end{figure}

\begin{figure}[h!]
\centering
%\psbox[xsize=0.4#1,ysize=0.2#1,rotate=r]
\psbox[xsize=9.0cm,rotate=r]{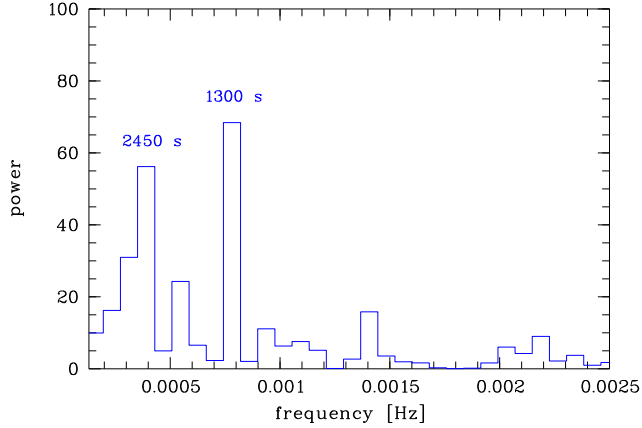}
\caption{Power spectrum for IRAS~13224--3809 in the 0.2-10 keV energy range.}
\label{fig_fft}
\end{figure}

To probe the quasi-periodic oscillation we used four independent tests. In
Fig. \ref{fig_lomb}, the strongest signal in the Lomb-Scargle periodogram is
2500 sec. The second strong signal at 1300 sec seems to be the higher order of the
signal at 2500 sec. The same result is given by the power spectrum of the
source in Fig. \ref{fig_fft}.

\begin{figure}[h!]
\centering
%\psbox[xsize=0.4#1,ysize=0.2#1,rotate=r]
\psbox[xsize=0.33#1,ysize=0.35#1,rotate=r]{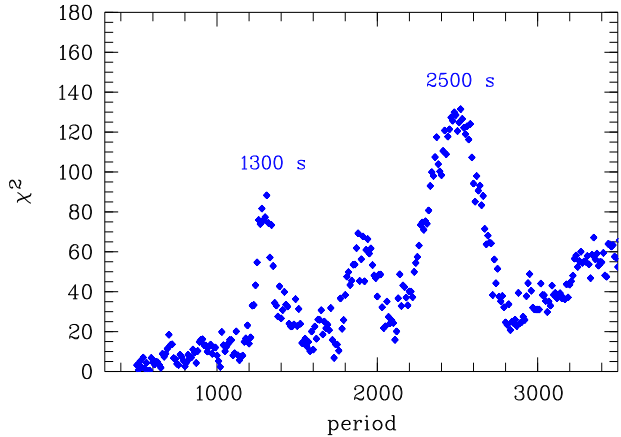}
\caption{Reduced $\chi^2$ versus the folding period for the 0.2-10 keV energy range.}
\label{fig_chi2}
\end{figure}

\begin{figure}[h!]
\centering
%\psbox[xsize=0.4#1,ysize=0.2#1,rotate=r]
\psbox[xsize=9.0cm,rotate=r]{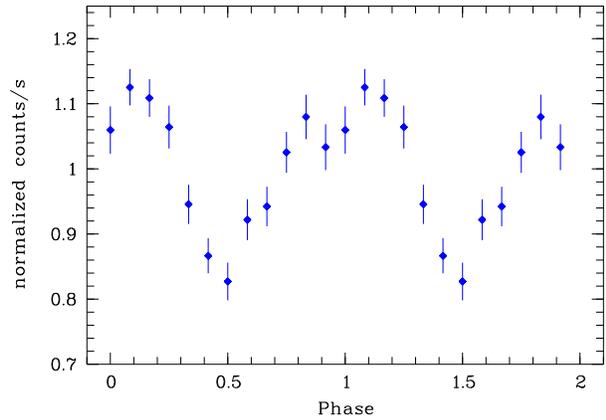}
\caption{The light curve of IRAS~13224--3809, folded with a period of 2380 s.}
\label{fig_flight}
\end{figure}

\begin{figure}[h!]
\centering
%\psbox[xsize=0.4#1,ysize=0.2#1,rotate=r]
\psbox[xsize=9.0cm,rotate=r]{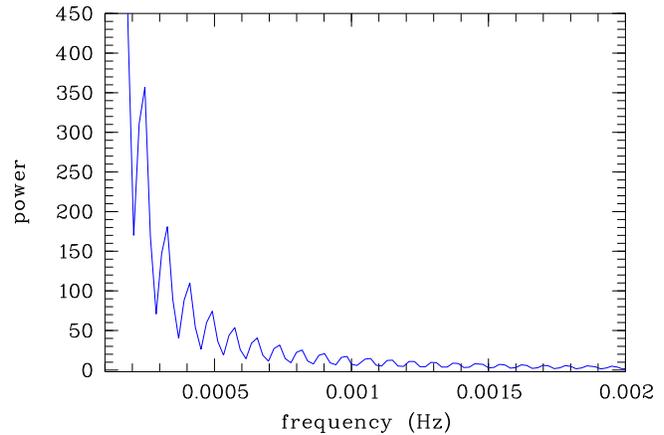}
\caption{Simulated red noise power spectrum of IRAS~13224--3809, normalized to
the real data set.}
\label{fig_rednoise}
\end{figure}

We have folded the light curve with different
periods, ranging from 500 to 3500 seconds, and have determined the
corresponding  $\chi^2$ value. The reduced $\chi^2$ versus the folding period
in Fig. \ref{fig_chi2} shows also the strongest peak at 2500 sec. 
The corresponding folded light curve is given in Fig. \ref{fig_flight} with the
best fitting period of 2400 sec and modulations of about 20 \%. 
We made simulations of light curves with the
observed time sequence and  phase randomized for a red noise $f^{-1}$ power
spectrum (see Fig. \ref{fig_rednoise}). The indication for the periodic signal
in the light curve of \iras (see the power spectrum in Fig. \ref{fig_fft}) are
in the same order compared to the power seen in the simulated power spectrum. Therefore
we need the 80 ksec \xmm  observation of the source to confirm the presence of
any quasi-peridic signal.

\vspace{-0.5cm}

\section{Discussion}
\label{bh_mass}

The quasi-periodic oscillations if confirmed by other X-ray observations, can
be caused by instabilities of the inner parts of the accretion disk, which can
lead to variations of the accretion rate. In this case the period is expected
to be of the order of the radial drift time scale ("instability time scale")
 
\[
\tau \sim \frac{\epsilon}{\alpha} \cdot \left(\frac{r}{R_S}\right)^\frac{3}{2}
\cdot \frac{R_S}{c} 
\]

with $\epsilon \sim 5-10$ and $\alpha \sim 0.1$ (Sunyaev, private communication).
For a period of 2500 sec and a mass of $\rm M_{BH}^{Edd} \sim 3 \cdot 10^6\;
M_{\odot}$ derived from the Eddington limit, we have estimated a radius of $\rm
r \approx 1.4\; R_S$, where the instabilities occur, which would be also
indicative for the presence of a Kerr black hole.
Absorption by Compton-thick matter can probably ruled out by an estimation
for the distance of the absorbing matter.

\begin{figure}[h]
\centering
\psbox[xsize=0.7#1,ysize=0.5#1]{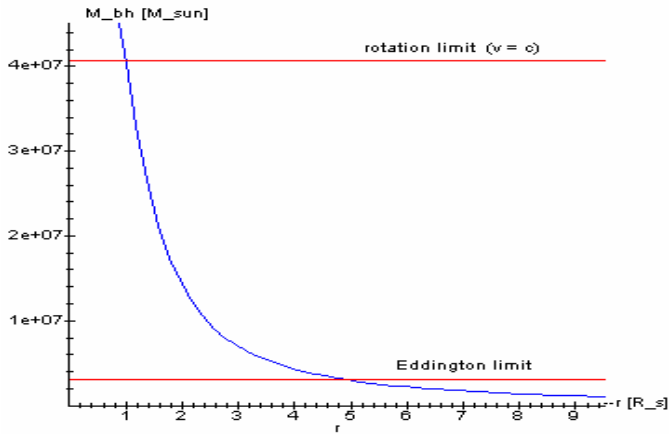}
\caption{Dependence of black hole mass to different rotation radii for
rotating Compton-thick matter (blue curve).  The radius is given in units of
the Schwarzschild radius and the black hole mass in units of solar mass.}
\label{fig_bh_mass}
\end{figure}

In Fig. \ref{fig_bh_mass} we show
the lower and upper value of the black hole mass for IRAS~13224--3809. The lower mass is given by
the Eddington limit ($\rm M_{BH}^{lower} \sim 3 \cdot 10^6\; M_{\odot}$). The
upper mass limit is estimated from material orbiting 
with $\rm v=c$ ( $\rm M_{BH}^{upper} \sim 4 \cdot 10^7\; M_{\odot}$). The blue
curve demonstrates the dependence of black hole mass to different rotation radii in units of
the Schwarzschild radius, assuming that rotating Compton-thick matter is causing the
X-ray periodicity (semi-relativistic calculation)\footnote{$\rm \frac{v}{c} =
\sqrt{\frac{r}{2(r-1)^2+r}}$; r in units of $\rm R_S$ and $\rm v = \frac{2\pi
\tilde{r}}{T}$; $\rm \tilde{r} = r \cdot R_S$}.

The putative Compton-thick
absorbing material has to be located at distances of $\rm \sim 5\;R_S$. This
seems to be inconsistent with neutral Compton-thick absorbing material, because
it is too close to the central black hole.

\section{Summary}
\label{discussion}

The precisely determined X-ray positon of the flaring source \iras (Boller et
al. 1997) is in good agreement with the optical position from the Hubble GSC of the
galaxy (within 2 arcsec). 
The light curve of \iras shows no huge flaring events as seen by Boller et al. (1997);
however, we found indications for a quasi-periodic signal in the light curve
of \iras with a period of 2500 sec. This quasi-periodic signal might be caused
by instabilities of inner parts of the accretion disk. To confirm and
to study the triggering processes of the putative periodic signal we need the
scheduled \xmm observation which will yield both timing and spectral
information of this source.

In addition, we found evidence for the presence of a Kerr black hole in
IRAS~13224--3809. The first hint is given by the change of luminosity over
time, which exceeds the maximum allowed value by accretion onto a Schwarzschild
black hole. The second one is the estimation of the distance of the accretion
disk instabilities to the black hole, resulting in the short distance of about
1.4 Schwarzschild radii. We estimate the black hole mass of \iras between $\rm
3\cdot10^6\;M_{\odot}$ and $\rm 4\cdot10^7\;M_{\odot}$.

\section*{References}

\re
Boller Th., Brandt W.N., Fink H., 1996 A\&A, 305, 53 

\re
Boller Th., Brandt W.N., Fabian A.C., Fink H.H., 1997 MNRAS, 289, 393

\re
Sunyaev R.A., 1973, Soviet Astron. AJ, 16, 941

\re
Sunyaev R.A., 2000, private communication

\label{last}

\end{document}